\documentclass[iop]{emulateapj}
\usepackage{gensymb}
\usepackage{wasysym}
\usepackage{float}
\usepackage{graphicx}
\usepackage{color}
\usepackage{url}
\usepackage[normalem]{ulem}

\newcommand{\arcname}{SGAS\,J143845.1$+$145407}
\newcommand{\clustername}{SDSS\,J1438$+$1454}
\newcommand{\arcshort}{SGAS\,J1438}
\newcommand{\zcluster}{0.237}
\newcommand{\zarcA}{0.816}

\newcommand{\HST}{{\it HST}}
\newcommand{\GALFIT}{\texttt{GALFIT}}
\newcommand{\Halpha}{H$\alpha$}
\newcommand{\Msun}{M$_{\odot}$}

\newcommand{\magtot}{$11.8^{+4.6}_{-2.4}$}
\newcommand{\maghigh}{10}
\newcommand{\maglow}{2}
\newcommand{\massc}{$6.0^{+0.3}_{-0.7}\times10^{12}$ \Msun}
\newcommand{\imageplanerms}{$0\farcs07$}

\slugcomment{ApJ in prep: draft date \today}

\shorttitle{SGAS\,J143845.1$+$145407: Morphology and Star Formation}
\shortauthors{Dunham et al.}

\begin{document}

\title{Lens Model and Source Reconstruction Reveal the Morphology and Star 
Formation Distribution in the Cool Spiral LIRG SGAS\,J143845.1$+$145407}

\author{Samuel J. Dunham\altaffilmark{1,2},
Keren Sharon\altaffilmark{1},
Michael K. Florian\altaffilmark{3},
Jane R. Rigby\altaffilmark{3},
Michael D. Gladders\altaffilmark{4,5},
Matthew B. Bayliss\altaffilmark{6},
H\aa kon Dahle\altaffilmark{7},
Traci L. Johnson\altaffilmark{1},
Katherine Murray\altaffilmark{1,8},
Katherine E. Whitaker\altaffilmark{9},
and Eva Wuyts\altaffilmark{10}}

\altaffiltext{1}{Department of Astronomy, The University of Michigan, 1085 South University Ave, Ann Arbor, MI 48109, USA}
\altaffiltext{2}{Vanderbilt University, Department of Astronomy, 6301 Stevenson Center Lane, Nashville TN, 37212}
\altaffiltext{3}{Observational Cosmology Lab, NASA Goddard Space Flight Center, 8800 Greenbelt Road, Greenbelt, MD 20771, USA}
\altaffiltext{4}{Kavli Institute for Cosmological Physics, University of Chicago, 5640 South Ellis Ave, Chicago, IL 60637, USA}
\altaffiltext{5}{Department of Astronomy and Astrophysics, University of Chicago, 5640 South Ellis Ave, Chicago, IL 60637, USA}
\altaffiltext{6}{MIT-Kavli Center for Astrophysics and Space Research, 77 Massachusetts Avenue, Cambridge, MA, 02139, USA}
\altaffiltext{7}{Institute of Theoretical Astrophysics, University of Oslo, P.O. Box 1029, Blindern, N-0315 Oslo, Norway}
\altaffiltext{8}{Space Telescope Science Institute, Baltimore, MD, USA}
\altaffiltext{9}{Department of Physics, University of Connecticut, Storrs, CT 06269, USA}
\altaffiltext{10}{Armen TeKort Antwerp, Belgium}

\begin{abstract}
We present \textit{Hubble Space Telescope} (\textit{HST}) imaging and grism spectroscopy
of a strongly lensed LIRG at $z=0.816$, SGAS 143845.1$+$145407, and use the magnification boost
of gravitational lensing to study the distribution of star formation throughout this galaxy.  Based
on the \HST\ imaging data, we create a lens model for this system; we compute the mass distribution
and magnification map of the $z=0.237$ foreground lens. We find that the magnification of the
lensed galaxy ranges between \maglow\ and \maghigh, with a total magnification (measured over all the
images of the source) of $\mu$=\magtot. We find that the total projected mass density within $\sim34$ kpc of the brightest
cluster galaxy is \massc. Using the lens model we create a source reconstruction for SGAS 143845.1$+$145407,
which paired with a faint detection of \Halpha\ in the grism spectroscopy, allows us to finally comment directly on
the distribution of star formation in a $z\sim1$ LIRG. We find widespread star formation across this galaxy,
in agreement with the current understanding of these objects. However, we note a deficit of \Halpha\ emission in the
nucleus of SGAS 143845.1$+$145407, likely due to dust extinction.
\end{abstract}


\section{Introduction}
Observations of luminous infrared galaxies (LIRGs, galaxies with total infrared luminosity (8-1000 $\mu$m) of $10^{11-12} L_{\odot}$ and a star formation rate (SFR) of $\mathtt{\sim}10-100\, M_{\odot}\ \text{yr}^{-1}$), which dominate star formation at $z\sim 1$, indicate that they evolve strongly
with redshift.

Spectroscopy and spectral energy distribution (SED)
measurements of LIRGs indicate that at $z=0$ they are more likely to
have warmer SEDs than their $z\sim1$
counterparts (Rowan-Robinson et al., 2004), and that the temperature of
the dust-reprocessed infrared radiation gets colder with increasing
redshift (Rowan-Robinson et al. 2005; Sajina et al. 2006; Symeonidis
et al. 2009; Elbaz et al., 2010; Hwang et al., 2010).
Several studies (e.g., Papovich et al., 2007; Rigby et al., 2008;
Farrah et al., 2008; Men{\'e}ndez-Delmestre et al., 2009) show that
the strength of the polycyclic aromatic hydrocarbon features increase
with redshift for galaxies with constant luminosity.
Morphologically, radio and submm interferometry show that the spatial extent of star formation in LIRGs and
ULIRGs (ultraluminous infrared galaxies--those with total infrared luminosity $>10^{12} L_{\odot}$ and a SFR exceeding $\mathtt{\sim}100 M_{\odot}\ \text{yr}^{-1}$) is considerably larger at $0.4<z<2.5$ than in the local universe (Rujopakarn et al. 2011).

These multiple lines of evidence therefore indicate that the mode of
star formation in LIRGs evolves significantly with redshift: At $z=0$,
star formation occurs on sub-galactic scales of hundreds of parsecs at
most, resulting in hot infrared SEDs and weak
aromatic features, while star formation at higher redshifts is
dominated by low SFRs in galaxy-wide
bursts, resulting in cooler SEDs and stronger aromatic features.

However, existing direct imaging studies lack the resolution required to distinguish between these
two star formation morphologies at high redshift.  At $z>0.4$, galaxies are unresolved by Spitzer and
Herschel, for example, and even with interferometry in the radio and submm, a beam size of $0\farcs3-0\farcs5$
is only a few times smaller than the size of the sources, making it impossible to robustly test whether the
distribution of star formation is different in $z\sim1$ LIRGs than it is in those at $z\sim0$.

This spatial resolution limitation can be overcome if we are able to take advantage of the magnifying power
of gravitational lensing.  When paired with the already-high resolution of telescopes like the \textit{Hubble Space
Telescope} (\textit{HST}), gravitational lensing enables observations on scales 10-100 times smaller than otherwise possible,
providing insights into the detailed physical properties of high redshift galaxies (e.g., Wuyts et al. 2012; Bayliss
et al. 2014; Bordoloi et al. 2016).  In fact, using gravitational lensing, Johnson et al. (2017a) recently measured
the sizes of individual star forming regions in a galaxy at $z=2.49$ and found some as small as
40 parsecs across, demonstrating the importance of this magnification boost to studying the details
of star formation in high redshift galaxies.  This means that in order to directly determine whether the mode of
star formation in LIRGs really does evolve between $z=1$ and $z=0$, the ideal target to observe
would be a strongly lensed galaxy at $z\sim1$ with properties typical of LIRGs at that redshift.

\arcname\ is such an object. First reported by Gladders et al. (2013), \arcname\ (hereafter \arcshort) is
a bright lensed galaxy behind the lensing cluster \clustername\ (Figure~1).
Using photometry in 18 bands from the blue
optical to $500\ \mu m$, and optical and near infrared spectroscopy, these authors determined that \arcshort\
is a cool LIRG undergoing dusty star formation. Other than being highly magnified, \arcshort\ is intrinsically a
typical LIRG at $z=0.816$.

We take advantage of this unique opportunity to study the details of star formation in a prototypical high-redshift LIRG.  
We present new \HST\ imaging data of \clustername\ in \S~\ref{S.data} and a lens
model based on new constraints from these data in \S~\ref{S.lensmodel}.  We discuss the implications of this lens
model in \S~\ref{S.results}. Using the lens model we produce a source plane reconstruction of \arcshort\ in
\S~\ref{S.source}, which we pair with \HST\ grism spectroscopy data in \S~\ref{S.SFdistribution} to constrain the spatial
extent of star formation in this $z=0.816$ LIRG. This allows us to directly test the current understanding of the evolution
of star formation in LIRGs from $z=1$ to $z=0$.

Where necessary, we assume a flat cosmology with $\Omega_{\Lambda}
= 0.7$, $\Omega_{m} =0.3$, and $H_0 = 70$  km s$^{-1}$
Mpc$^{-1}$. In this cosmology $1\arcsec$ corresponds to $3.758$ kpc at 
the cluster redshift, $z=$\zcluster, and $7.559$ kpc at the source redshift, $z=$\zarcA. 
Magnitudes are reported in the AB system.

\section{Data}\label{S.data}
\clustername\ was observed by \HST\ Cycle 21 program GO13437 (PI: Rigby)
during three orbits. Imaging with the Advanced Camera for Surveys (ACS) was
executed over one orbit on GMT 2014 March 5, in F814W (600 s) and
F606W (780 s). 
Wide Field Camera 3 (WFC3) grism spectroscopy was
conducted in two different roll angles, one on GMT 2014 March 5 and
one on GMT 2014 May 30, using the G141 grism (2406 s  each).
F140W imaging frame was taken at each roll angle 
(285 s each), giving a total imaging time in this band of 570 s.
The ACS images were taken with a 3-point line dither using
a 1000$\times$1000 pixel subraster to manage buffer dumps, resulting in
three frames per filter. Since the charge transfer efficiency (CTE) of
the ACS detector is decreasing, post-observation image corrections
were applied to individual exposures using the pixel-based empirical
CTE correction software provided by Space Telescope Science Institute (STScI).
Individual frames were then combined using AstroDrizzle (Gonzaga at al. 2012) 
with a pixel scale of $0\farcs03$ pixel$^{-1}$, and a drop size of 0.5 (WFC3) and 0.8
(ACS), following Sharon et al. (2014).  
All the images were aligned and registered to the same pixel frame as
the F140W image.

The slitless spectroscopy data were reduced using the Grizli
pipeline\footnote[1]{https://github.com/gbrammer/grizli}.  
Each visit was processed
separately rather than being stacked because the data were taken with
different roll angles.  The two-dimensional spectra of each of the four
images of \arcshort\ are significantly contaminated by 
light from the brightest cluster galaxy (BCG) and other bright galaxies.
To minimize the effects of the light from these
objects we deviate from the standard Grizli procedure by modeling the contamination 
from the galaxies that directly affect the 
spectra of \arcshort\ separately from the rest of the field, using \GALFIT\ (Peng et al. 2002, 2014).
We find that for the purpose of assigning light to individual 
objects in areas where light from multiple galaxies overlaps,
the \GALFIT\ modeling is better than the Source Extractor segmentation maps
that are part of the standard Grizli pipeline.  Because of this we use the following alternative
method (G. Brammer, private communication): We truncate the \GALFIT\ models 
when the flux from an object falls to less than 0.001 of
the flux of the central pixel. 
The
\GALFIT\ models were then subtracted from the direct images, and the
initial contamination models were computed for the remaining objects.  The initial
contamination models for the two sets of galaxies were combined before
the final refinement of the models was conducted.  This two-part
modeling process resulted in two-dimensional spectral extractions with
substantially less contamination from bright nearby galaxies than the
standard Grizli pipeline.  However, as we will discuss in \S~\ref{S.SFdistribution},
contamination was still significant. 

\begin{figure}
\centering
\includegraphics[scale = 0.47]{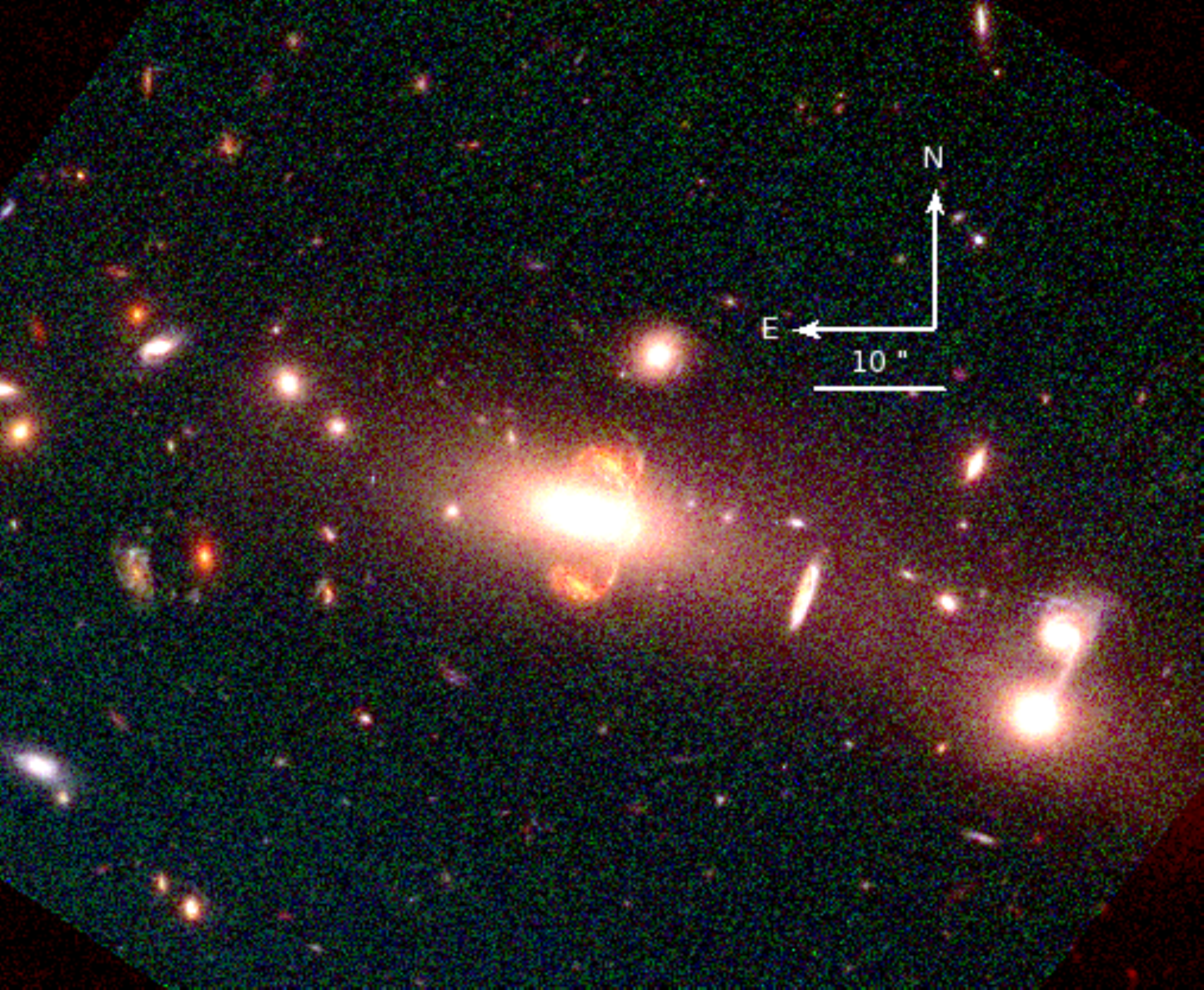}
\caption{Composite color \HST\ image of SDSS J1438+1454 in F140W (red), 
  F814W (green) and F606W (blue). The
  \HST\ imaging data are consistent with this being a low-mass
  cluster, dominated by the BCG and a few cluster member
  galaxies. In these data two of the multiple images of \arcshort\ are easily
  identified north and south of the BCG. There are also two partial images that
  lie to the east and west of the BCG, but they are obscured by the intracluster light.}
\label{fig.original}
\end{figure}

\begin{figure}
\centering
\includegraphics[scale=0.44]{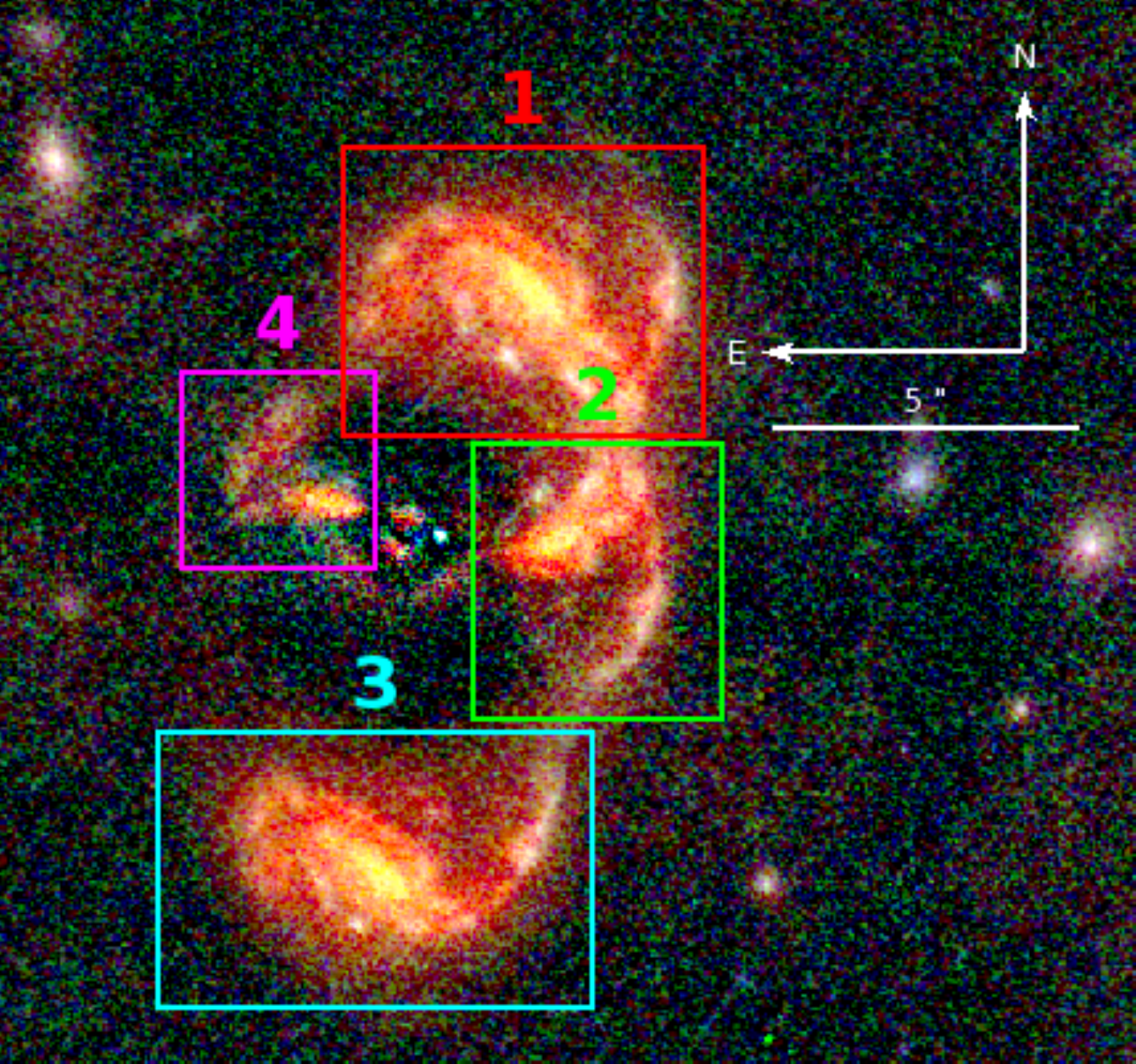}
\caption{A zoom-in view of \arcshort; the light of the BCG was
  modeled with \GALFIT\ (Peng et al. 2002, 2014) and subtracted from the
  imaging data. The filters are the same as in
  Figure~\ref{fig.original}. We mark each of the unique images of
  \arcshort\ with boxes. Images 1 and 3 (red and cyan boxes, respectively) are
  of the same parity and approximately the same magnification. Images
  2 and 4 (green and magenta boxes, respectively) are partial
  images. We note a residual unresolved blue clump close to the center
  of the BGC that is probably due to emission in the lens galaxy and
  not to lensing. }
\label{fig.nobcg}
\end{figure}

\section{Strong Lens Model}\label{S.lensmodel}
To analyze the images of the lensed galaxy we model the light of the
foreground BCG with a multi-component S\'{e}rsic
profile using \GALFIT, and subtract the model 
from the imaging data in each band. The result
(Figure~\ref{fig.nobcg}) reveals the images of the lensed galaxy that
are buried in the light of the foreground BCG.

The high resolution \HST\ data confirm the lensing interpretation of
Gladders et al. (2013), which was based on lower-resolution imaging in 18 bands, spanning
0.5 to 500 $\mu m$, from Gemini, Magellan, \textit{Spitzer} Space Telescope, 
\textit{Wide-field Infrared Survey Explorer} (\textit{WISE}), and the \textit{Herschel} Space Observatory.
The strong lensing potential of the foreground
cluster causes the appearance of four images of \arcshort. Two of these
are complete images appearing to the north and south of the BCG
(labeled 1 and 3 in Figure~\ref{fig.nobcg}, respectively); and two are partial images
appearing west and east of the BCG
(labeled 2 and 4, respectively).
In each of the images we identify a number of distinct emission knots,
most prominent in F606W and F814W.
 
\begin{figure}
\centering
\includegraphics[scale=0.27]{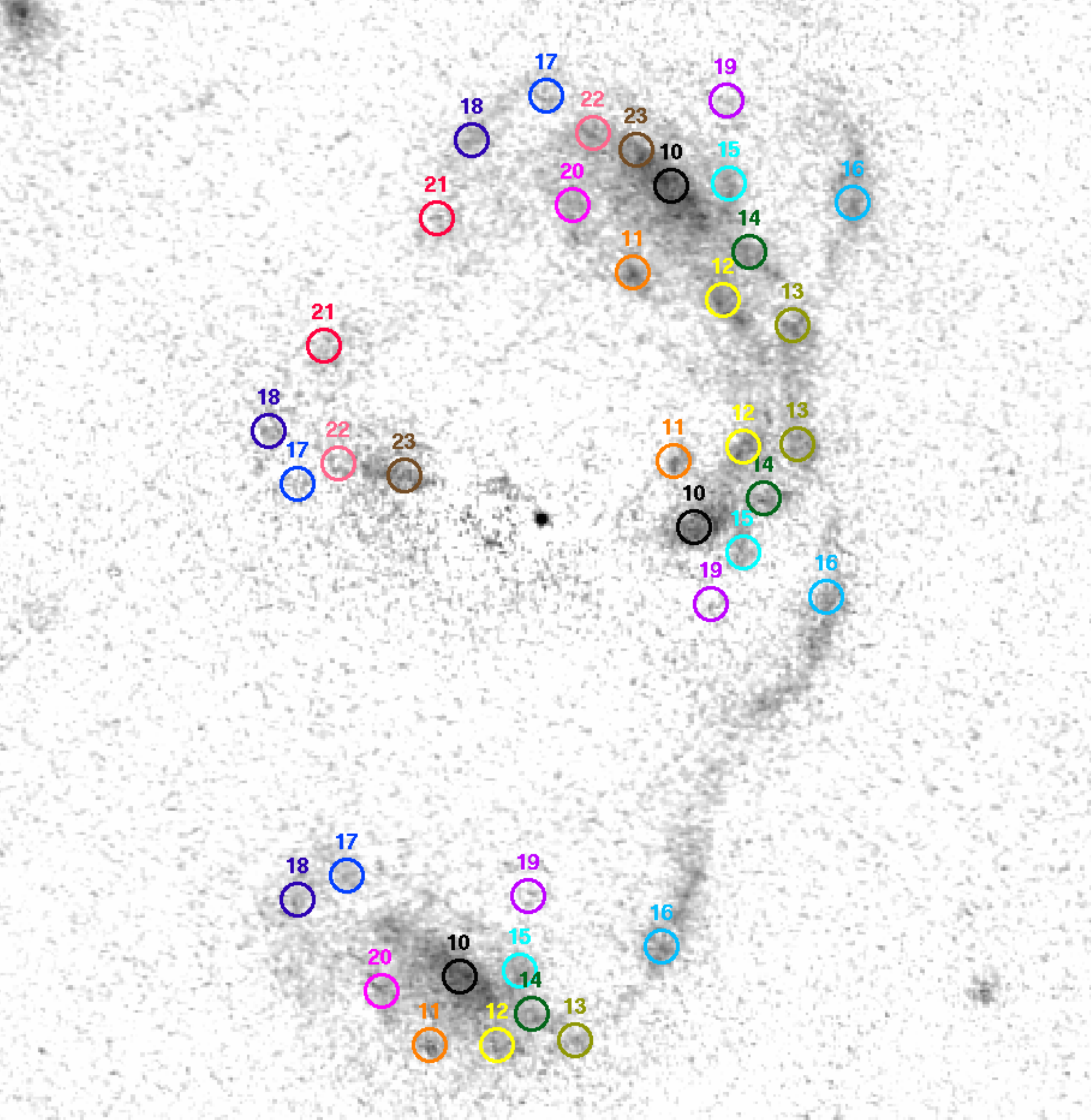}
\caption{The emission knots in \arcshort\ that were used to constrain
  the lens model are labeled over the F814W image. See also Table~\ref{tab.constraints}.}
\label{fig.constraints}
\end{figure}

We matched the multiply-imaged knots between instances of the lensed
galaxy by considering the location of the knots within the galaxy,
lensing parity, color, magnification, and brightness. Several emission
knots are easily distinguishable, and thus are used to constrain the
lens model (Figure~\ref{fig.constraints}). Including the bulge of the galaxy,
a total of $38$ emission knots were used as positional
constraints. Table~\ref{tab.constraints} lists the IDs and coordinates
of the emission knots that were used in the lens model.

The strong lens model was computed with the public software \texttt{Lenstool}
(Jullo et al. 2007), which uses a Bayesian Markov Chain Monte Carlo
(MCMC) method to explore the parameter space. The final minimization was done
in the image plane by requiring the smallest scatter between the
predicted and observed positions of images of each emission knot.
We modeled the lens plane as a linear combination of several halos,
representing the cluster, the BCG, and cluster-member galaxies.
Each halo is parameterized as a pseudo isothermal ellipsoidal mass
distribution (PIEMD; Jullo et al. 2007), with the following
parameters: position $x$, $y$; ellipticity $e\equiv(a^2-b^2)(a^2+b^2)^{-1}$, where $a$
and $b$ are the semi-major and semi-minor axes, respectively; position angle $\theta$; core and cut
radii $r_{core}$ and $r_{cut}$, respectively; and velocity dispersion
$\sigma$. All the parameters of the cluster halo were allowed to vary with the
exception of the cut radius, which cannot be constrained by the strong lensing
evidence and was therefore fixed at $r_{cut}$=1000 kpc. We verify that this assumption 
does not affect the results of this work.
In addition to the cluster halo we placed a halo that represents the
central galaxy; its position was fixed on its observed
coordinates, its cut radius was fixed at $r_{cut}$=20 kpc, and all other parameters were left free.

Cluster-member galaxies were identified from the \HST\ photometry by
their F814W-F140W color in a color-magnitude diagram (Gladders et al. (2000)). The positions,
ellipticities, and position angles of each galaxy were fixed to their observed values,
and the PIEMD profile parameters were scaled to their F140W luminosity
using scaling relations, following Limousin et al. (2005).

Our model is in general agreement with that of Gladders et al. (2013).
In particular, we find that the position angle, core radius, 
and velocity dispersions of the cluster and of the BCG are within the
statistical uncertainties of the respective models. We note, however, 
that due to the smaller number of lensing
constraints that were available to Gladders et al. (2013), they had simplified 
their lens model by stronger assumptions on the extent to which the mass 
distribution follows the light distribution. 
They limit the center of the cluster to not be further than 
$8\farcs01^{+2.01}_{-3.90}$ from the BCG, in agreement with our findings of $4\farcs88$.
The best-fit model has an image plane rms scatter of \imageplanerms. Table 1 lists the rms of each image.

\begin{deluxetable}{llllll}
\tablecaption{Lensing Constraints}
\tablehead{ %
\colhead{Image ID} & %
\colhead{R.A.(J2000)} & %
\colhead{Decl.(J2000)} & %
\colhead{rms ($^{\prime\prime}$)} & %
\colhead{$\left|\mu\right|$}}
\startdata
10.1 & 219.68704 & 14.904572  & 0.05 & 3.2$_{-0.7}^{+0.6}$ \\
10.2 & 219.68696 & 14.903430  & 0.04 & 1.8$_{-0.4}^{+1.6}$ \\
10.3 & 219.68777 & 14.901929  & 0.04 & 2.4$_{-0.5}^{+0.4}$ \\
\hline
11.1 & 219.68717 & 14.904282 & 0.03 & 5.4$_{-1.0}^{+2.4}$ \\
11.2 & 219.68703 & 14.903652 & 0.11 & 2.8$_{-0.5}^{+2.5}$ \\
11.3 & 219.68787 & 14.901701 & 0.07 & 2.1$_{-0.4}^{+0.2}$ \\
\hline
12.1 & 219.68686 & 14.904191 & 0.01 & 6.4$_{-1.8}^{+2.0}$ \\
12.2 & 219.68679 & 14.903699 & 0.09 & 4.5$_{-0.8}^{+2.5}$ \\
12.3 & 219.68764 & 14.901701 & 0.06 & 2.2$_{-0.4}^{+0.2}$ \\
\hline
13.1 & 219.68662 & 14.904108 & 0.05 & 8.0$_{-2.9}^{+1.4}$ \\
13.2 & 219.68660 & 14.903708 & 0.05 & 6.4$_{-0.8}^{+3.2}$ \\
13.3 & 219.68737 & 14.901716 & 0.05 & 2.3$_{-0.5}^{+0.2}$ \\
\hline
14.1 & 219.68677 & 14.904353 & 0.02 & 4.2$_{-1.1}^{+0.8}$ \\
14.2 & 219.68672 & 14.903530 & 0.04 & 2.9$_{-0.8}^{+1.5}$ \\
14.3 & 219.68752 & 14.901806 & 0.02 & 2.4$_{-0.5}^{+0.3}$ \\
\hline
15.1 & 219.68684 & 14.904579 & 0.04 & 3.1$_{-0.7}^{+0.5}$ \\
15.2 & 219.68679 & 14.903349 & 0.08 & 2.1$_{-0.6}^{+1.1}$ \\
15.3 & 219.68756 & 14.901952 & 0.06 & 2.6$_{-0.6}^{+0.4}$ \\
\hline
16.1 & 219.68641 & 14.904516 & 0.03 & 3.1$_{-0.8}^{+0.2}$ \\
16.2 & 219.68650 & 14.903197 & 0.02 & 2.9$_{-1.2}^{+0.6}$ \\
16.3 & 219.68707 & 14.902029 & 0.05 & 3.1$_{-0.8}^{+0.5}$ \\
\hline
17.1 & 219.68747 & 14.904875 & 0.08 & 2.7$_{-0.5}^{+0.4}$ \\
17.3 & 219.68816 & 14.902269 & 0.06 & 2.9$_{-0.6}^{+0.7}$ \\
17.4 & 219.68833 & 14.903578 & 0.04 & 1.7$_{-0.2}^{+1.4}$ \\
\hline
18.1 & 219.68773 & 14.904724 & 0.14 & 3.3$_{-0.7}^{+0.8}$ \\
18.3 & 219.68833 & 14.902189 & 0.07 & 2.6$_{-0.5}^{+0.5}$ \\
18.4 & 219.68843 & 14.903754 & 0.02 & 2.1$_{-0.4}^{+1.3}$ \\
\hline
19.1 & 219.68685 & 14.904857 & 0.06 & 2.5$_{-0.5}^{+0.3}$ \\
19.2 & 219.68690 & 14.903173 & 0.06 & 2.1$_{-0.5}^{+1.2}$ \\
19.3 & 219.68753 & 14.902201 & 0.04 & 3.2$_{-0.8}^{+0.9}$ \\
\hline
20.1 & 219.68738 & 14.904508 & 0.02 & 3.8$_{-0.7}^{+1.3}$ \\
20.3 & 219.68804 & 14.901883 & 0.02 & 2.3$_{-0.5}^{+0.4}$ \\
\hline
21.1 & 219.68785 & 14.904465 & 0.08 & 5.6$_{-1.1}^{+2.7}$ \\
21.4 & 219.68824 & 14.904039 & 0.06 & 4.6$_{-1.1}^{+1.7}$ \\
\hline
22.1 & 219.68731 & 14.904751 & 0.03 & 2.9$_{-0.6}^{+0.6}$ \\
22.4 & 219.68819 & 14.903645 & 0.02 & 1.4$_{-0.1}^{+1.7}$ \\
\hline
23.1 & 219.68716 & 14.904693 & 0.18 & 3.0$_{-0.6}^{+0.5}$ \\
23.4 & 219.68796 & 14.903606 & 0.16 & 1.6$_{-0.2}^{+4.7}$
\enddata
\label{tab.constraints}
\tablecomments{The lensing constraints used in the model of \arcshort; see Figure~\ref{fig.constraints}.
The image plane rms is given in arcseconds for each image, and their magnifications are listed for the best-fit model. The 1$\sigma$ uncertainties are computed from the steps in the MCMC sampling.}
\end{deluxetable}

\section{Implications of the Lens Model}\label{S.results}
The best-fit model is obtained by an MCMC sampling of the parameter
space, with a total of ten free parameters (six for the cluster halo
and four for the BCG). Table~\ref{tab.results} shows the best-fit
parameters and their $1\sigma$ uncertainties, as obtained from the
MCMC sampling. 

The critical curves and magnification contours of the best-fit model for a source at $z$=\zarcA\
are shown in Figure~\ref{fig.magnific}.

\subsection{The Cluster Mass}
We find that the lens plane is dominated by an elongated group-scale
halo, centered $4\farcs88$ west of the central galaxy,  in the
direction of the second and third brightest galaxies in this group (see Figure \ref{fig.mass} and Figure \ref{fig.original}). 

Strong lensing is highly sensitive to the total projected mass
density within the strong lensing region, i.e. out to
projected radii where lensed images are observed, or approximately the
Einstein radius of the lens (e.g., Meneghetti et al. 2010; Johnson \& Sharon 2016).
We therefore report the projected mass density enclosed within   
R=$9\arcsec$($\approx34$ kpc) centered on the
BCG, M=\massc.
The $1\sigma$ statistical uncertainties are estimated by computing 1000 lens models from 
the MCMC sets of parameters and calculating the mass distribution for each.

The total mass
of the cluster would be best measured by weak lensing or other mass
proxies, and is beyond the scope of this work. Nevertheless,
extrapolation of the strong lensing mass out to 500 kpc ($\sim
135\arcsec$) yields a total enclosed mass of M=$1\times10^{14}$ \Msun. 
We note that this number has a large statistical and
systematic uncertainty, due to
the inability of strong lensing alone to constrain the mass outside of
the strong lensing region.

\subsection{Magnification}
\begin{figure}
\centerline{\includegraphics[width=\linewidth]{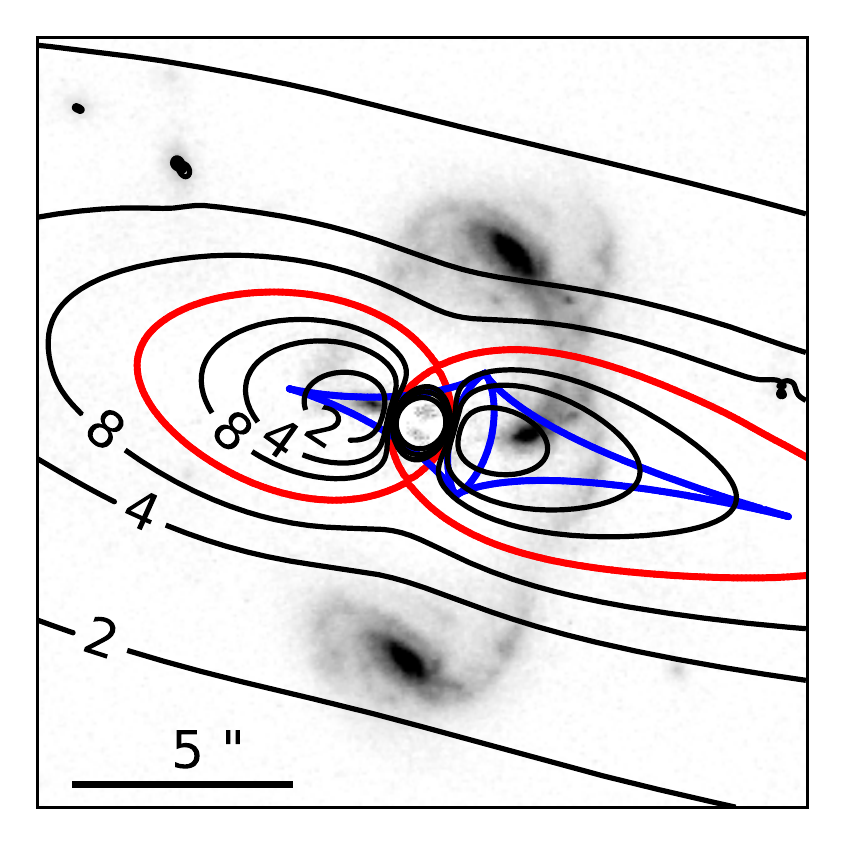}}
\caption{Contours of the absolute value of magnification as calculated
  from the best-fit lens model, for a source at $z$=\zarcA. In red we plot the critical curve,
  i.e. regions in the image plane with formally infinite
  magnification. In blue we plot the caustic, which is the projection of the critical curve onto the source plane. In this figure north is up and east is to the left.}
\label{fig.magnific}
\end{figure}

\begin{figure}
\includegraphics[width=\linewidth]{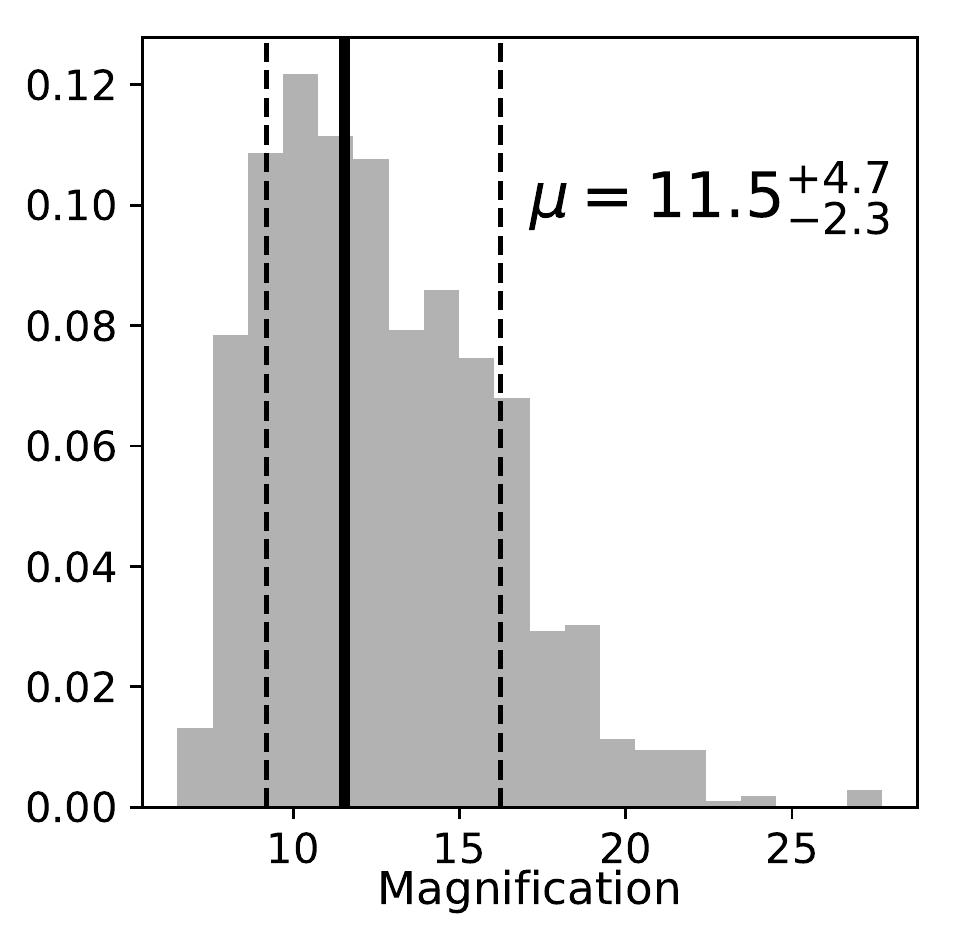}
\caption{The total magnification of \arcshort\ is computed as the
  ratio of the total image plane area to the source plane area of the multiple images (see
  text). The solid line denotes the value from the best-fit model and the dashed lines 
  mark the $1\sigma$ confidence intervals, computed from the MCMC chain.}
\label{fig.maghist}
\end{figure}

\begin{figure}
\centerline{\includegraphics[width=\linewidth]{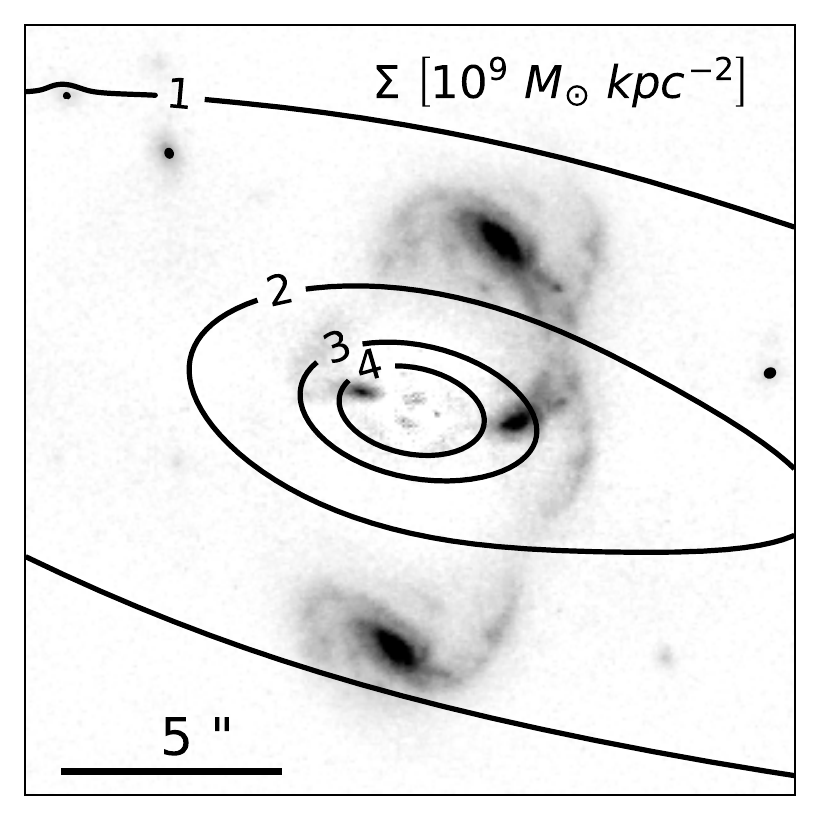}}
\caption{The surface mass density contours from the best-fit model
  overplotted on the F140W image with the BCG removed. The mass density distribution is
  dominated by the BCG and elongated in the direction of the next
  brightest galaxy. In this figure north is up and east is to the left.}
\label{fig.mass}
\end{figure}

Figure~\ref{fig.magnific} shows the magnification contours from the
best-fit model. The magnification changes by about a factor of three
across each image.  The total magnification, 
\magtot,
is computed as follows: We first determine the surface brightness of the background by choosing a patch 
of the image that is mostly sky, and compute the brightness distribution of pixels in this region. We then choose an 
isophote that is 5$\sigma$ brighter than the background--this defines a set of polygons in the image plane. 
We calculate the areas of these polygons
using Green's theorem. The polygons are then ray-traced to the source
plane using the lensing equation and the deflection matrix from the lens model (see \S~\ref{S.source}). The areas are re-calculated in the source plane, again using Green's theorem, and the magnification is then computed as the
ratio of the total area in the image plane to the total area in the
source plane. The procedure was repeated for 1000 models from the MCMC
steps in order to derive the magnification uncertainties (see Figure~\ref{fig.maghist}). 
This procedure results in a non-weighted average magnification of the source galaxy.

The magnification of individual emission knots is less affected by the magnification gradient, given their size. We list the magnification from the best-fit model at the position of each of the emission clumps in Table~\ref{tab.constraints}; the uncertainties are calculated from the 1000 MCMC models.

\section{Source Plane Reconstruction}\label{S.source}
Following the procedure in Sharon et al. (2012, 2014) we
ray-trace the image plane pixels from each image of the source through
the lensing equation, $\vec\beta = \vec\theta - d_{ls}/d_s\,
\vec\alpha(\vec\theta)$, where $\vec\beta$  and $\vec\theta$ are the source and image
positions, respectively, $d_{ls}/d_s$ is the ratio of the angular
diameter distances from the lens to the source and from the observer to
the source, and $\vec\alpha( \vec\theta )$ is the deflection matrix of the best-fit
model. The coordinates and shape of each pixel are translated to the
source plane while conserving surface brightness. 
Figure~\ref{fig.source} shows the reconstruction of the source plane
from each one of the four images of \arcshort. Since images 2 and 4
are partial images they do not form a complete image of the source
galaxy in the source plane as do images 1 and 3. In the source plane,
the positions of the ray traced emission knots have a mean scatter of $0\farcs033$,
indicating the agreement between the source reconstructions of the different images.

Unlensed, the spatial extent of the galaxy in the source plane
is $\sim5$ square arcseconds, which translates to $\sim287$ kpc$^2$ at the source
redshift, $z$=\zarcA. The spatial extent of the source is defined by taking the area of the images, 
ray-tracing them back to the source-plane, and re-calculating the area in the source-plane.
The source plane morphology is similar to the
lensed morphology of the galaxy, indicating that the lensing potential
does not introduce significant distortion. 
The lensing magnification is generally a combination of an isotropic magnification component
due to the local projected mass density, and an anisotropic component due to shear. 
Low distortion indicates that in this location the lens produces little shear, resulting in nearly isotropic magnification.

\section{Morphology and Distribution of Star Formation}\label{S.SFdistribution}
\begin{figure*}
\centering
\includegraphics[scale=0.32]{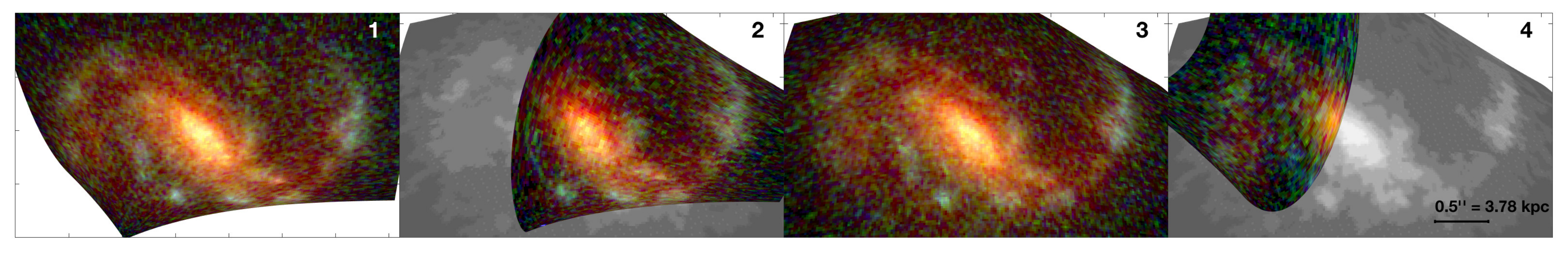}
\caption{
  \arcshort\ is reconstructed from each of its
  four images by ray-tracing the image-plane pixels through the
  deflection of the best-fit lens model. Images 1 and 3 form complete images of the source,
  while images 2 and 4 are partial images, due to their location with
  respect to the source plane caustic. The grayscale background in
  panels 2 and 4 is to guide the eye to the extent of the full image
  of the source, and is replicated from image 3. For context, the source plane scale is given
as a horizontal bar.}
\label{fig.source}
\end{figure*}

The source plane reconstructions (Figure~\ref{fig.source}) reveal \arcshort\ to be a large, two-arm, grand design spiral galaxy. This was suggested by the image plane morphology but is even more evident given the source plane reconstructions. The red color of the core of this galaxy and the shape of the SED published in Gladders et al. 2013 suggests that the nuclear emission is extremely obscured by dust.

In contrast, the spiral arms contain many distinct patches of emission with colors much bluer than the core, suggesting that there is widespread recent star formation in the arms that is largely unobscured by dust.  At z=0.816, \Halpha\ emission, a tracer of ongoing star formation, falls at a wavelength that does not lie within any of the three broadband filters used to observe \arcshort.  It is, however, within the wavelength range observable in our HST WFC3/IR grism spectroscopy.

\begin{figure*}
\centering
\includegraphics[scale=0.32]{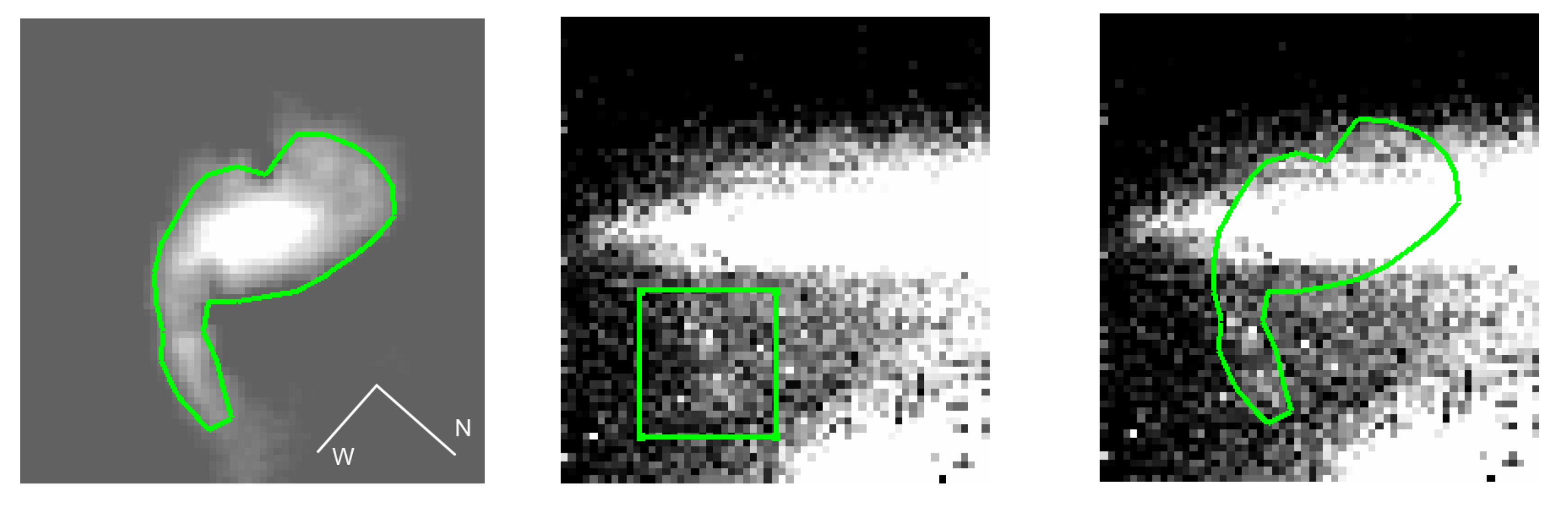}
\caption{
Left: Direct image of \arcshort\ image 3 in the F140W filter, with an isophote drawn as a reference for the object's shape.  Center: The 2D spectral extraction from HST WFC3/IR G141 grism spectroscopy data, showing a very faint \Halpha\ detection (marked by the box) below the bright nuclear continuum emission.  Right: The isophote from the leftmost panel on top of the 2D grism spectral extraction in the center panel, with the isophote placed so that the flux weighted center of the direct image would lie at the wavelength of \Halpha\ emission at the redshift of \arcshort.  Note that the slight flux excess noted in the center panel lies inside the the spiral arm, aligned closely with the emission in the direct image.
}
\label{fig.grism2D}
\end{figure*}

\Halpha\ emission is not detected in the nucleus of \arcshort\ in our grism observations.  Furthermore, because of the orientation of the galaxy the brightest emission in the spiral arms of images 1 and 3 is closely aligned with the much stronger continuum emission from the nucleus in the direction of 
dispersion, causing the nuclear emission to drown out the signal in the spiral arms.  Unfortunately, grism observations at other angles would have caused the BCG contamination to overwhelm the spectra of images 1 and 3.  That same contamination, at the angles used in our grism program, makes it impossible to access the spectra of images 2 and 4.  However, in image 3, the northwestern spiral arm is successfully separated from the nucleus, albeit with moderate to high contamination from the BCG.  A hint of \Halpha\ emission is visible, as shown in Figure~\ref{fig.grism2D}. Comparing an isophote of the direct image (left panel of Figure~\ref{fig.grism2D}) with the shape of the faint emission in the 2D grism spectrum at the wavelength of \Halpha\ at z=0.816 (center panel) shows very faint \Halpha\ emission coincident with the large blue clumps in the spiral arm seen on the right of panel~3 in Figure~\ref{fig.source}. A suggestion of \Halpha\ emission from the southeastern portion of image 3 is also visible above the spectrum of the nucleus, but it is even less clear.

Because of the contamination from the BCG it is difficult to estimate the strength of the \Halpha\ emission in the spiral arms. The fact that the \Halpha\ emission aligns spatially with one of the blue regions in the broadband imaging bolsters the interpretation that it and the many other similarly colored regions throughout the spiral arms are regions of ongoing star-formation. Unfortunately, with these data it is difficult to comment on the star formation in the nucleus because it is so severely obscured by dust. Interestingly, Gladders et al. 2013 reported a detection of \Halpha\ in the nucleus, which we fail to detect here.  With \HST-quality imaging, however, we see that there are blue clumps---likely \Halpha\ emitters---to the south and southeast of the nucleus and hints of others to the north and northwest that were included in the slit of the Gladders et al. 2013 spectroscopy.  We now believe that the nuclear \Halpha\ emission reported in that paper is actually emission from these blue clumps that were not identifiable  in extant ground-based imaging at the time.

The finding that star forming regions are large (the largest such regions are on the kiloparsec scale according to the reconstructed source image) and distributed throughout the galaxy with a morphology consistent with a grand-design spiral lends support to the interpretation that star formation occurs on galaxy-wide scales in z$\sim$1 LIRGs rather than in small (hundreds of parsecs or smaller), localized regions like in z$\sim$0 LIRGs. One caveat to this apparent galaxy-scale star formation is the deficit of observed \Halpha\ emission in the core, which probably is not a deficit of star formation but likely reflects enhanced obscuration at that location. Unfortunately, the extant long-wavelength imaging on \arcshort\ lack the spatial resolution to clearly demonstrate extensive obscured star formation in the core. Understanding whether or not the core is actively forming stars would be useful for understanding what physical conditions lead to this galaxy-wide mode of star formation; but in either case we have \textit{directly} shown that the star formation in this z$\sim$1 LIRG is indeed occurring at much larger spatial scales than in z$\sim$0 LIRGs, as was previously expected but not directly observed.

\section{Discussion}
We find that the mass of the lens is consistent with a small group of
galaxies (e.g., Han et al. 2015), in line with the cluster richness
reported in Gladders et. al. (2013) of
$N^{weighted}_{gals}=9.734$. The lens model favors a displacement of
the center of the cluster halo from the BCG, at
$4\farcs88$ in the direction of the second and third brightest
galaxies in this structure. A small offset between the BCG and the
center of mass of the group or cluster is expected (e.g., George et
al. 2012), and consistent with the light 
distribution of the cluster (Figure~\ref{fig.original}). Our model is in good agreement with the observed
velocity dispersion of the BCG reported in 
\cite{gladders13}, $\sigma_v=318\pm111$ km s$^{-1}$. 

The magnification of \arcshort\ allows us to study star
formation in the source galaxy in detail that would be unattainable
without lensing. The typical lensing magnification of individual emission
knots, a factor of 3-5,  enables studies of
individual star forming regions in a galaxy at $z$=\zarcA\ with a high
signal-to-noise ratio. The angular
size of the source galaxy, unmagnified, is $2\farcs53$ ($\approx19$
kpc). In good conditions, ground-based resolution of $0\farcs6$
 would allow at best up to four non-overlapping
resolution elements, dominated by the bulge. With the lensing magnification
boost even ground-based observations can resolve more than ten regions in
this galaxy and thus resolve its two-arm structure. 

\HST\ imaging has revealed the structure of this galaxy to be a large, two-arm, grand design spiral with a very red core and bluer clumps with sizes up to the kiloparsec scale scattered through its spiral arms.  \HST/WFC3 grism spectroscopy of all the images of the galaxy were severely contaminated and dominated by light of the foreground BCG; 
high-resolution integral field spectroscopy would overcome this issue. 
Nevertheless, the grism data confirm that \Halpha\ emission is present in these clumps, though no such emission is detected in the core.  The widespread star formation in the spiral arms supports previous interpretations of the spatially unresolved spectra of z$\sim$1 LIRGs, namely that such objects are likely to have large scale, nearly galaxy-wide star formation, in contrast to their counterparts in the local universe where star formation can be intense but occurs on smaller scales.

Based on observations made with the NASA/ESA Hubble Space Telescope, obtained at the Space Telescope
Science Institute, which is operated by the Association of Universities for Research in Astronomy, Inc., under
NASA contract NAS 5-26555. These observations are associated with program \#14420.

\acknowledgments
We thank the anonymous referee for their constructive feedback that significantly improved this paper.
Support for program GO-14420 was provided by NASA through a grant from
the Space Telescope Science Institute. 
This work makes use of the Matlab Astronomy Package (Ofek 2014).

\begin{deluxetable}{lll}
\tablewidth{230pt} 
\tablehead{\colhead{Parameter} & \colhead{Cluster} & \colhead{BCG}}
\startdata
R.A. [$^{\prime\prime}$] & $4.34^{+1.06}_{-2.86}$ & [$0$] \\
Decl. [$^{\prime\prime}$] & $-1.08^{+0.73}_{-0.34}$ & [$0$] \\
e & $0.87^{+0.07}_{-0.13}$ & $0.47^{+0.08}_{-0.36}$ \\
$\theta$ [\degree] & $165^{+2}_{-1}$ & $168^{+28}_{-29}$ \\
$r_{core}$ [kpc] & $31^{+39}_{-31}$ & $1.58^{+2.87}_{-0.67}$ \\
$\sigma$ [km/s] & $606^{+3}_{-96}$ & $281^{+39}_{-51}$ \\
$r_{cut}$ [kpc]     &  [$1000$] & [$20$] 
\enddata
\label{tab.results}
\tablecomments{Best-fit values from the image plane optimization. The coordinates of the BCG 
are fixed at their observed location, (R.A., Decl.) = (219.68768, 14.903482). The position of the cluster 
halo is measured relative to the BCG. The position angle $\theta$ is measured north of west, and the 
ellipticity of the projected mass density is $e=(a^2-b^2)/(a^2+b^2)$, where $a$ and $b$ are the 
semi-major and semi-minor axes, respectively. The core radius of the cluster was not well-constrained by the model, and so the uncertainties quoted here are the priors set for the MCMC optimization.}
\end{deluxetable}

\acknowledgments


\begin{thebibliography}{0}%
\makeatletter
\providecommand \@ifxundefined [1]{%
 \@ifx{#1\undefined}
}%
\providecommand \@ifnum [1]{%
 \ifnum #1\expandafter \@firstoftwo
 \else \expandafter \@secondoftwo
 \fi
}%
\providecommand \@ifx [1]{%
 \ifx #1\expandafter \@firstoftwo
 \else \expandafter \@secondoftwo
 \fi
}%
\providecommand \natexlab [1]{#1}%
\providecommand \enquote  [1]{``#1''}%
\providecommand \bibnamefont  [1]{#1}%
\providecommand \bibfnamefont [1]{#1}%
\providecommand \citenamefont [1]{#1}%
\providecommand \href@noop [0]{\@secondoftwo}%
\providecommand \href [0]{\begingroup \@sanitize@url \@href}%
\providecommand \@href[1]{\@@startlink{#1}\@@href}%
\providecommand \@@href[1]{\endgroup#1\@@endlink}%
\providecommand \@sanitize@url [0]{\catcode `\\12\catcode `\$12\catcode
  `\&12\catcode `\#12\catcode `\^12\catcode `\_12\catcode `\%12\relax}%
\providecommand \@@startlink[1]{}%
\providecommand \@@endlink[0]{}%
\providecommand \url  [0]{\begingroup\@sanitize@url \@url }%
\providecommand \@url [1]{\endgroup\@href {#1}{\urlprefix }}%
\providecommand \urlprefix  [0]{URL }%
\providecommand \Eprint [0]{\href }%
\providecommand \doibase [0]{http://dx.doi.org/}%
\providecommand \selectlanguage [0]{\@gobble}%
\providecommand \bibinfo  [0]{\@secondoftwo}%
\providecommand \bibfield  [0]{\@secondoftwo}%
\providecommand \translation [1]{[#1]}%
\providecommand \BibitemOpen [0]{}%
\providecommand \bibitemStop [0]{}%
\providecommand \bibitemNoStop [0]{.\EOS\space}%
\providecommand \EOS [0]{\spacefactor3000\relax}%
\providecommand \BibitemShut  [1]{\csname bibitem#1\endcsname}%
\let\auto@bib@innerbib\@empty
\end{thebibliography}%


\begin{thebibliography}{}
\bibitem[Bayliss et al.(2014)]{2014ApJ...790..144B} Bayliss, M.~B., Rigby, J.~R., Sharon, K., et al.\ 2014, \apj, 790, 144 
\bibitem[Bordoloi et al.(2016)]{2016MNRAS.458.1891B} Bordoloi, R., Rigby, J.~R., Tumlinson, J., et al.\ 2016, \mnras, 458, 1891 
\bibitem[Elbaz et al.(2010)]{elbaz10} Elbaz, D., Hwang, H.~S., Magnelli, B., et al.\ 2010, A\&A, 518, L29
\bibitem[Farrah et al.(2008)]{farrah08} Farrah, D., Lonsdale, C.~J., Weedman, D.~W., et al.\ 2008, ApJ, 677, 957
\bibitem[George et al. (2012)]{george12} George, M.~R., Leauthaud, A., Bundy, K., et al.\ 2012, ApJ, 757, 2
\bibitem[Gladders et al.(2000)]{Gladders2000} Gladders, M.~D., \& Yee, H.~K.~C.\ 2000, AJ, 120, 2148.
\bibitem[Gladders et al. (2013)]{gladders13} Gladders, M.~D., Rigby, J.~R., Sharon, K., et al.\ 2013, ApJ, 764, 177
\bibitem[Gonzaga \& et al.(2012)]{2012hstp.book.....G} Gonzaga, S., \& et al.\ 2012, Hubble Space Telescope  Primer for Cycle 21
\bibitem[Gonzaga \& et al.(2012)]{2012drzp.book.....G} Gonzaga, S., \& et al.\ 2012, The DrizzlePac Handbook, HST Data Handbook
\bibitem[Hwang et al.(2010)]{hwang10} Hwang, H.~S., Elbaz, D., Magdis, G., et al.\ 2010, MNRAS, 409, 75
\bibitem[Han et al. (2015)] {han15} Han, J., Eke, V.~R., Frenk, C.~S., et al. 2015, MNRAS, 446, 1356
\bibitem[Johnson \& Sharon(2016)]{2016ApJ...832...82J} Johnson, T.~L., \& Sharon, K.\ 2016, \apj, 832, 82 
\bibitem[Johnson et al.(2017)]{2017ApJ...843L..21J} Johnson, T.~L., Rigby, J.~R., Sharon, K., et al.\ 2017, \apjl, 843, L21 
\bibitem[Jullo et al.(2007)]{jul07} Jullo, E., Kneib, J.-P., Limousin,  M., El{\'{\i}}asd{\'o}ttir, {\'A}., Marshall, P.~J., \& Verdugo, T.  2007, New Journal of Physics, 9, 447
\bibitem[Limousin et al.(2005)]{2005MNRAS.356..309L} Limousin, M., Kneib, J.-P., \& Natarajan, P.\ 2005, \mnras, 356, 309 2001, \apj, 563, 9
\bibitem[Meneghetti et al.(2010)]{2010A&A...514A..93M} Meneghetti, M., Rasia, E., Merten, J., et al.\ 2010, \aap, 514, A93 
\bibitem[Men{\'e}ndez-Delmestre et al.(2009)]{karin09}  Men{\'e}ndez-Delmestre, K., Blain, A.~W., Smail, I., et al.\ 2009, ApJ, 699, 667
\bibitem[Ofek(2014)]{2014ascl.soft07005O} Ofek, E.~O.\ 2014, Astrophysics Source Code Library, ascl:1407.005 
\bibitem[Papovich et al.(2007)]{papovich07} Papovich, C., Rudnick, G., Le Floc'h, E., et al.\ 2007, ApJ, 668, 45
\bibitem[Peng et al.(2002)]{2002AJ....124..266P} Peng, C.~Y., Ho, L.~C., Impey, C.~D., \& Rix, H.-W.\ 2002, \aj, 124, 266 
\bibitem[Peng et al.(2010)]{2010AJ....139.2097P} Peng, C.~Y., Ho, L.~C., Impey, C.~D., \& Rix, H.-W.\ 2010, \aj, 139, 2097 
\bibitem[Rigby et al.(2008)]{rigby08} Rigby, J.~R., Marcillac, D., Egami, E. et al.\ 2008, ApJ, 675, 262
\bibitem[Rowan-Robinson et al.(2004)]{rowan-robinson04} Rowan-Robinson, M., Lari, C., Perez-Fournon, I., et al.\ 2004, MNRAS, 351, 1290
\bibitem[Rowan-Robinson et al.(2005)]{rowan-robinson05} Rowan-Robinson, M., Babbedge, T., Surace, J., et al.\ 2005, AJ, 129, 1183
\bibitem[Rujopakarn et al.(2011)]{ruj11} Rujopakarn, W., Rieke, G.~H., Eisenstein, D.~J., \& Juneau, S.\ 2011, ApJ, 726, 93
\bibitem[Sajina et al.(2006)]{sajina06} Sajina, A., Scott, D., Dennefeld, M., et al.\ 2006, MNRAS, 369, 939
\bibitem[Sharon et al.(2012)]{ker12} Sharon, K., Gladders, M.~D., Rigby, J.~R., et al.\ 2012, ApJ, 746, 161
\bibitem[Sharon et al.(2014)]{2014ApJ...795...50S} Sharon, K., Gladders, M.~D., Rigby, J.~R., et al.\ 2014, \apj, 795, 50 
\bibitem[Symeonidis et al.(2009)]{symeonidis} Symeonidis, M., Page, M.~J., Seymour, N., et al.\ 2009, MNRAS, 397, 1728
\bibitem[Wuyts et al.(2012)]{evamassZ} Wuyts, E., Rigby, J.~R., Sharon, K., \& Gladders, M.~D. 2012, ApJ, 755, 73.
\end{thebibliography}
\end{document}